\documentclass[a4paper,12pt]{article}
\usepackage{amsmath}
\usepackage{amsfonts}
\usepackage[dvips]{graphicx}
\usepackage[bookmarksnumbered=true,breaklinks=true]{hyperref}
\usepackage{a4}
\usepackage{parskip}
\newcommand{\ri}{{\rm i}} 
\newcommand{\vare}{\varepsilon} 
\newcommand{\xid}{\xi \cdot \partial} 

\newcommand{\dpad}[2]{{\displaystyle{\frac{\partial #1}{\partial #2}}}}
\newcommand{\dfud}[2]{{\displaystyle{\frac{\delta #1}{\delta #2}}}}
\newcommand{\ddfrac}[2]{{\displaystyle{\frac{#1}{#2}}}}
\newcommand{\dsum}[2]{\displaystyle{\sum_{#1}^{#2}}}   
\newcommand{\dint}{\displaystyle{\int}}
\newcommand{\eq}{\begin{equation}}
\newcommand{\eqn}[1]{\label{#1}\end{equation}}
\newcommand{\ba}{\begin{array}}
\newcommand{\ea}{\end{array}}

\newcommand{\lp}{\left(}\newcommand{\rp}{\right)}

\newcommand{\hd}{{\hat\delta}}

\newcommand{\slav}{{\cal S}}
\newcommand{\BB}{{\cal B}}
\newcommand{\St}{S_{\rm tot}}
\newcommand{\bS}{{\bar S}}

\newcommand{\hAstar}{{{\hat A}^*}{}}

\newcommand{\inv}[1]{\frac{1}{#1}}
\newcommand{\starco}[2]{\left[#1\stackrel{\star}{,}#2\right]}
\newcommand{\staraco}[2]{\left\{#1\stackrel{\star}{,}#2\right\}}
\newcommand{\co}[2]{\left[#1,#2\right]}
\newcommand{\var}[2]{\frac{\d #1}{\d #2}}
\newcommand{\vvar}[3]{\frac{\d^2 #1}{\d #2\d #3}}
\newcommand{\diff}[2]{\frac{\partial #1}{\partial #2}}

\newcommand{\intx}{\int d^4x}

\newcommand{\Gam}{\Gamma}
\newcommand{\Act}{S}
\newcommand{\e}{\epsilon}
\newcommand{\s}{\sigma}
\renewcommand{\d}{\delta}
\renewcommand{\l}{\lambda}
\renewcommand{\th}{\theta}
\renewcommand{\k}{\tilde{k}}

\newcommand{\bc}{{\bar{c}}}
\newcommand{\uim}{UV/IR mixing }
\newcommand{\nc}{noncommutative }
\newcommand{\eth}[1]{\frac{\e_{#1}}{\th}}
\newcommand{\tri}{\triangle}
\newcommand{\dd}{\hat\delta}
\newcommand{\eid}{\vare \cdot \partial}
\newcommand{\nid}{n \cdot \partial}
\title{\begin{flushright}  {\normalsize LYCEN 2006-05}\end{flushright}\vspace{2ex}\bf A Vector Supersymmetry \\
in Noncommutative $U(1)$ Gauge Theory \\
With the Slavnov Term}
\author{\\[-0.3cm]\Large 
Daniel N.~Blaschke\footnote{work supported by ``Fonds zur F\"orderung der 
Wissenschaftlichen Forschung'' (FWF) under contract P15015-N08.}~, 
Fran\c cois Gieres\footnotemark[3]~, 
Olivier Piguet\footnote{work supported  
in part by the Conselho Nacional 
de Desenvolvimento Cient\'{\i}fico e  
Tecnol\'{o}gico CNPq -- Brazil.} \\[1.5mm] \Large
and  Manfred Schweda\footnotemark[1]
}
\date{}
\graphicspath{{figures/}}
\begin{document}
\maketitle
\begin{center}
\renewcommand{\thefootnote}{\fnsymbol{footnote}}
\vspace{-0.3cm}\footnotemark[1]Institute for Theoretical Physics, 
Vienna University of Technology\\
Wiedner Hauptstrasse 8-10, A-1040 Vienna (Austria)\\[0.3cm]
\footnotemark[3]Institut de Physique Nucl\'eaire,
Universit\'e Claude Bernard (Lyon 1), \\
4 rue Enrico Fermi, F - 69622 - Villeurbanne (France)\\[0.3cm]
\footnotemark[2]Departamento de F\'{\i}sica, CCE,
Universidade Federal do Esp\'{\i}rito Santo (UFES), \\
Av. Fernando Ferrari, 514, BR-29075-910 - Vit\'oria - ES (Brasil)\\[0.5cm]
\ttfamily{E-mail: blaschke@hep.itp.tuwien.ac.at, gieres@ipnl.in2p3.fr, opiguet@yahoo.com, mschweda@tph.tuwien.ac.at}
\vspace{0.5cm}
\end{center}
\begin{abstract}
We consider noncommutative $U(1)$ gauge theory with the additional term,
involving a scalar field $\lambda$, introduced by Slavnov in order to
cure the infrared problem. We show that this theory, 
with an appropriate space-like axial gauge-fixing, exhibits a 
linear vector supersymmetry similar to the one present 
in the $2$-dimensional $BF$ model. 
This vector supersymmetry implies that 
all loop corrections 
are independent of the $\lambda AA$-vertex and thereby explains 
why Slavnov found a finite model for the same gauge-fixing.
\end{abstract}

\newpage 

\tableofcontents

\section{Introduction}\label{sect1}
It is well known that  \nc quantum field theories (NCQFT's)
realized through the Weyl-Moyal  $\star$ product
suffer in general 
from the problem of UV/IR mixing~\cite{Filk:1996}. This implies 
that one has to deal 
with IR singularities for vanishing external momentum. 
In order to get rid of the \uim in noncommutative gauge field 
theories (NCGFT's) with $U(1)$ gauge group, 
Slavnov~\cite{Slavnov:2003,Slavnov:2004}  introduced an 
additional term of the following form 
into the action:
\begin{align}\label{sl-term}
\frac{1}{2}
\intx \, \l \star\th^{\mu\nu}F_{\mu\nu}
\, .
\end{align}
Here, $\l$ represents a new \emph{dynamical} 
 or \emph{``quantum''}  multiplier field 
and the constant antisymmetric tensor $(\th^{\mu\nu})$
describes the noncommutativity of space-time 
coordinates: $[x^\mu,x^\nu]$ $=$ 
$\ri \th^{\mu\nu}$.
As a consequence of this so-called 
{\em Slavnov term},  
the photon propagator becomes transversal 
with respect to the momentum 
$\k^\mu=\th^{\mu\nu}k_\nu$. Thereby,  
insertions of the (gauge independent) IR singular parts 
of the one-loop polarization 
tensor~\cite{Blaschke:2005b}
\begin{align}
\Pi^{\mu\nu}_{\text{IR}}(k)
=\frac{2g^2}{\pi^2} \, \frac{\k^\mu\k^\nu}{(\k^2)^2}
\, , 
\end{align}
are expected to vanish in higher-order loop calculations. 
For an axial gauge-fixing with 
$( n^\mu ) =(0,1,0,0)$ 
this result actually holds~\cite{Slavnov:2004}. 
However, for a covariant gauge-fixing, new problems arise 
due to the fact that one has new Feynman rules 
including a $\l$-propagator, a mixed $\l$-photon-propagator 
and a corresponding vertex --- see reference~\cite{Blaschke:2005c}
for a detailed discussion.

In this paper, we present a new approach by identifying 
the Slavnov term (\ref{sl-term}) with a topological term. 
In order to preserve the 
unitarity of the $S$-matrix~\cite{Susskind:2000},
we assume $\th^{\mu\nu}$ to be space-like,
i.e. $\th^{0i}=0$ in  suitable space-time 
coordinates. 
Furthermore, we can choose the spatial  coordinates 
in such a way that the only nonvanishing components 
of the $\theta$-tensor are $\th^{12}=-\th^{21}=\th$. 
Thus, the components $\th^{ij}$ with $i,j \in \{ 1, 2 \}$ 
can be written as $\th^{ij}=\th\e^{ij}$, where 
$\e^{ij}$ is the two-dimensional Levi-Civita 
symbol\footnote{We have 
$\e_{ij}\e^{kl} = \d_i^k\d_j^l-\d_i^l\d_j^k $.
}. 
The Slavnov term (\ref{sl-term}) then reads as 
$\frac{\th}{2}
\intx \, \l \star 
 \e^{ij} F_{ij}$
so that it resembles the action for a 
$2$-dimensional $BF$ model with Abelian gauge 
group~\cite{Schweda:1996}
\begin{align}\label{bf-2dim}
\Act_{\text{BF}}=\inv{2}\int d^2x \, 
\phi \, \e^{ij} F_{ij} \, .
\end{align}
The latter model represents a 
topological quantum field theory
and it is well known that such theories exhibit remarkable ultraviolet finiteness properties at the quantum level. 
In particular, the $3$-dimensional Chern-Simons theory and the $BF$ models in arbitrary space-time dimension 
represent fully finite quantum field theories. 
Their perturbative finiteness relies on the existence 
of a linear vector supersymmetry (VSUSY for short)
which is generated by a set of fermionic charges forming a Lorentz-vector~\cite{dgs, Piguet:1995}. Together with the scalar fermionic charge
of the BRST symmetry, they form a superalgebra of Wess-Zumino type, 
i.e. a graded algebra which closes on-shell on space-time translations. More precisely, one has the following graded commutation relations between the BRST operator $s$ and the operator $\d_\mu$ describing VSUSY: 
\begin{align}
\left\{s,\d_\mu\right\} \Phi=\partial_\mu\Phi
\; + \; 
\text{contact terms.}
\end{align}
Here,  $\Phi$ collectively denotes the basic fields 
appearing in the topological model under consideration 
and {\em contact terms} are expressions which vanish if the 
equations of  motion are used.
In this context, the axial gauge plays a special role since
the topological field theories mentioned above are characterized, in this gauge, by the complete absence of radiative corrections at the loop level.

We note that the \nc $2$-dimensional $BF$ model
is characterized, at least in the Lorentz gauge, by a VSUSY of the same form 
as in the commutative case~\cite{blasi:2005}.

The present paper is organized as follows.
In sections \ref{vector-susy}, 3 and  \ref{ST-identities} we discuss the symmetries of $U(1)$-NCGFT with Slavnov term along the lines of topological models with an axial gauge-fixing.
In section~\ref{loop-consequences}, we then elaborate on the consequences of these symmetries for higher-order loop calculations
and in particular we show that the VSUSY infers the absence 
of IR divergences (which was previously pointed out by 
Slavnov~\cite{Slavnov:2003,Slavnov:2004}).

\section{Symmetries of NCGFT with Slavnov term in 
the axial gauge}\label{vector-susy}
\subsection{Action}

The $U(1)$ gauge field action with Slavnov term and 
with an axial gauge-fixing~\cite{Blaschke:2005c} is 
given by
\begin{equation}
\label{action}
S= S_{{\rm inv}} + S_{{\rm gf}}
\, , \quad {\rm with} 
\ \left\{
\begin{array}{l}
 S_{{\rm inv}} =
\intx \ ( -\frac{1}{4}F_{\mu\nu}\star F^{\mu\nu}
+ 
\frac{1}{2} \l \star \, \th^{\mu\nu}F_{\mu\nu} ) 
\\
\\
 S_{{\rm gf}} =
 \intx \ ( B\star n^{\mu}A_{\mu}-\bc\star n^{\mu}D_{\mu}c )
\, , 
\end{array}
\right.
\end{equation}
where
\begin{eqnarray}
\label{defFD}
F_{\mu\nu}
 \!\!\! &=& \!\!\!
\partial_{\mu}A_{\nu}-\partial_{\nu}A_{\mu}-
\ri g\starco{A_{\mu}}{A_{\nu}}
\,,
\nonumber 
\\
D_{\mu} c  \!\!\! &=& \!\!\! \partial_{\mu} c
-\ri g\starco{A_{\mu}}{c}
\, .
\end{eqnarray}
  With  
$\th^{12}=-\th^{21}=\th$ as the only nonvanishing components 
of the 
  $\theta$ - tensor, 
the Slavnov term reduces to  $\frac{\theta}{2}
  \int d^4x \, 
\l \star \e^{ij} F_{ij} $,
i.e. (\ref{bf-2dim}) written as an integral over $4$-dimensional 
noncommutative space. 
  The axial gauge-fixing vector $n^\mu$ appearing in $S_{{\rm gf}}$
will be chosen to lie in the plane of noncommuting coordinates,
i.e. the plane $(x^1,x^2)$, hence $n^0=n^3=0$.   
We will see below that this allows us to find a VSUSY which is analogous
to the one characterizing 
the $2$-dimensional noncommutative $BF$ model. 

\subsection{Notation}

In order to distinguish the 
$x^1,x^2$-components  
from the other ones, we will use the following notation: 
Greek indices $\mu,\nu,\rho,\s \in \{ 0,1,2,3 \} $
correspond to the $4$-dimensional 
space-time,  Latin indices $i,j,k,l \in \{ 1,2 \} $ 
label  the $x^1,x^2$-components and capital Latin indices 
$I,J,K,L \in \{ 0,3 \}$ label the $x^0,x^3$-components.

For the particular 
choices of the axial gauge-fixing vector $(n_\mu)$ 
and the deformation matrix that we specified above, 
the action (\ref{action}) reads as
\begin{equation}\label{action-2dim}
\Act = \int d^4x
\ ( -\frac{1}{4}F_{\mu\nu}\star F^{\mu\nu}
+\frac{\th}{2}\l\star\e^{ij}F_{ij}
+B\star n^{i}A_{i}-\bc\star n^{i}D_{i}c )
\, .
\end{equation}
It is worthwhile recalling that the star product is associative 
and that it has the trace property
\[
\int d^4x \, f \star g = \int d^4x \, f \cdot g
= \int d^4x \, g \star f
\, , 
\]
henceforth we can perform cyclic permutations under the integral:
\begin{equation}
\int d^4x \, f \star g \star h = 
\int d^4x \, h \star f \star g = 
\int d^4x \, g \star h \star f 
\, .
\end{equation}
This property will often be used in the following. 

In order to simplify the notation, we will not spell out 
the star product symbol in the sequel:
{\bf all products between fields (or functions of fields) 
are understood to be star products}.
Furthermore, we assume that the algebra of fields is graded 
by the ghost-number. 
Accordingly, {\bf all commutators are considered to be graded}
with respect to this degree, 
e.g. $\frac{1}{2} {[c , c ]}$ stands for 
$\frac{1}{2} \staraco{c}{c} = c \star c$
and ${[ A_{\mu} , c ]}$ stands for  
$\starco{A_{\mu}}{c}$ $=$ $A_\mu\star c - c\star A_\mu$.

\subsection{Symmetries}

The action functional (\ref{action-2dim})  
is invariant 
under the {\em BRST transformations}
\begin{eqnarray}
sA_\mu \!\!\! &=& \!\!\! D_\mu c \, , 
\qquad  s \bar c  = B \, , 
\nonumber \\
 s\l  \!\!\! &=& \!\!\! - \ri g \, [\l, c ] \, , 
\qquad 
 sB = 0 \, , 
\label{BRS}
\\
sc  \!\!\! &=& \!\!\!   
\frac{\ri g}{2}  \, [c , c ] \, , 
\nonumber 
\end{eqnarray}
which are nilpotent, i.e. 
$s^2\Phi=0$ for $\Phi \in  \{ A_{\mu},\l,c,\bc, B \}$.
The functional (\ref{action-2dim}) is also invariant 
under the following {\em VSUSY transformations}
which are labeled by a vector index $i \in \{ 1, 2 \}$
and which only involve the  $x^1,x^2$-components of the fields:
\begin{eqnarray}
\label{susy}
\d_iA_J \!\!\! &=& \!\!\! 0 \, , 
\qquad \qquad 
\d_ic=A_i \, ,
\nonumber\\
\d_iA_j  \!\!\! &=& \!\!\! 0 \, , 
\qquad \qquad 
\d_i\bc=0 \, ,
\\
\d_i\l  \!\!\! &=& \!\!\! \frac{ \e_{ij} }{\th} n^j\bc \, ,
\qquad 
\d_iB=\partial_i\bc \, .
\nonumber
\end{eqnarray}
The noteworthy feature of these transformations is that they relate
the invariant and the gauge-fixing parts of the action (\ref{action-2dim}). Since the operator $\d_i$ lowers the ghost-number by one unit, it represents  an antiderivation (very much like the BRST operator $s$ which raises the  ghost-number by one unit).
Note that it is only the interplay of appropriate choices 
for  $\theta^{\mu\nu}$  
and  $n^\mu$  which leads to the existence of the VSUSY. The crucial point is the choice of the vector $n^\mu$ lying in the plane of noncommuting coordinates.

The invariance of the action functional (\ref{action-2dim})  
under the transformations (\ref{susy}) 
is described by the {\em Ward identity}
\begin{align}
\label{ward-susy}
\mathcal{W}_i\Act 
\equiv \intx \,( 
\partial_i \bc \, \var{\Act}{B} 
+A_i \, \var{\Act}{c} 
+ \frac{\e_{ij}}{\th}  n^j \bc \, \var{\Act}{\l} )
=0
\, .
\end{align}
For later reference, we determine the equations of motion 
associated to the action (\ref{action-2dim}). They are given by
$\var{\Act}{\Phi}=0$  
where $\Phi$ denotes a generic field. One finds that  
\begin{subequations}\label{eom}
\begin{align}
\var{\Act}{c}&=-n^iD_i\bc
\, , \qquad \qquad 
\var{\Act}{\bc} = -n^iD_ic \, , 
\label{eom1}\\
\var{\Act}{A_i}&= D_\mu F^{\mu i} +\th\e^{ij}D_j\l
+ n^iB- \ri gn^i [ \bc , c ]
\, , 
\label{eom3a}\\
\var{\Act}{A_I}&=  D_\mu F^{\mu I}
\, , \qquad \qquad \qquad 
\var{\Act}{\l} =\frac{\th}{2}\e^{ij}F_{ij}=\th F_{12}
\, , 
\label{eom-lambda} \\
\var{\Act}{B} & =n^i A_i
\, .
\label{gf}
\end{align}
\end{subequations}
The equation of motion for $\l$ implements the Slavnov condition 
$\e^{ij}F_{ij}=0$, 
i.e. the vanishing of the third
component of the magnetic field: $B_3=0$. 
The equation of motion for $B$ implements an axial gauge 
condition $n^i A_i=0$.

From equations (\ref{BRS}) and (\ref{susy}), we can deduce the  
graded commutation relations of the  BRST and VSUSY 
transformations. By using expressions (\ref{eom}),
the results can be cast into the following form: 
\begin{equation}\label{algebre}
{[ s, s ]} \, \Phi= {[ \d_i, \d_j ]}  \, \Phi=0 
\hspace{1cm} 
\text{for}\ \Phi \in \{A_{\mu},\l,c,\bc, B \}
\end{equation}
and 
\begin{subequations}\label{algebra}
\begin{align}
&{[s, \d_i ]} \,  \Phi =\partial_i \Phi 
\qquad \qquad \quad \ \ {\rm for} \  \Phi \in \{c,\bc, B \} \, ,  
 \label{algebra11}\\
& {[s, \d_i ]} \,  A_J= \partial_iA_J - F_{iJ} 
\,,\label{algebra1} \\
& {[ s, \d_i ]} \,  A_j=\partial_iA_j
-\frac{\e_{ij}}{\th}\var{\Act}{\l}
\, , \label{algebra14}\\
&{[ s , \d_i ]}  \, \l =\partial_i\l+\frac{\e_{ij}}{\th}\var{\Act}{A_j}
- \frac{1}{\th^2} D_i\var{\Act}{\l}
- \frac{\e_{ij}}{\th}D_KF^{Kj}.\label{algebra5}
\end{align}
\end{subequations}
Since contact terms appear in the graded commutators,  
the algebra can only close on-shell. 
Note that, apart from the translations,
the commutators (\ref{algebra1}) and (\ref{algebra5})
involve some extra contributions 
which are not related to equations of motion. 
One can readily verify that these terms represent a {\it new symmetry} 
of the action (\ref{action}) defined by the following 
field variations:
\begin{eqnarray}
\label{newsymmetry}
\hd_i A_J \!\!\!& = & \!\!\!  -F_{iJ}\,,
\qquad 
\hd_i\l = -\ddfrac{\e_{ij} }{\th} D_K F^{Kj}
\,, \\
\hd_i\Phi   \!\!\!& = & \!\!\! 0 \qquad \quad 
\mbox{for all other fields} \, .
\nonumber 
\end{eqnarray}
Concerning the proof, we only note that the transformations (\ref{newsymmetry})
and the Bianchi identity imply 
\[
\hd_i F_{JK} = -D_i F_{JK} \, , 
\qquad 
\hd_i F_{jK} = -D_i F_{jK} -D_K F_{ij} 
\, .
\]
Note, that the operator $\hd_i$ does not change the ghost-number.

Together with the BRST transformations, the 
VSUSY and the translations in the $(x^1,x^2)$-plane, 
\eq
 \delta_i^{({\rm transl})} \Phi = \partial_i\Phi\,,
\eqn{translation}
this new symmetry forms an algebra which actually closes on-shell: 
the translations commute with all transformations and 
\begin{equation}
 \label{brst-new}
\left.
\begin{array}{l}
\ \ {[s, s]} \, \Phi = 
{[s, \hd_j]} \, \Phi = 0 \\
{[ \d_i, \d_j ]} \, \Phi
= {[ \d_i , \hd_j ]} \, \Phi = 0
\end{array} \right\} 
\quad  \mbox{for all fields} \ \Phi
\, , \end{equation}
\begin{subequations}\label{algebraa}
\begin{align}
&{[s,\d_i]}\, \Phi =\partial_i  \Phi + \hd_i \Phi  
\qquad \quad {\rm for} \ \Phi \in \{ A_J, c, \bc, B \} 
\,, \label{algebra2'}\\
&{[s , \d_i]} \, A_j = \partial_i A_j+ \hd_i A_j
-\frac{\e_{ij}}{\th} \var{\Act}{\l}
\, ,\label{algebra11'}\\
&{[s, \d_i]} \, \l =\partial_i \l+ \hd_i\l+\frac{\e_{ij}}{\th}\var{\Act}{A_j}
- \frac{1}{\th^2} D_i\var{\Act}{\l}
\,,
\label{algebra5'}
\end{align}
\end{subequations}
and 
\begin{eqnarray}
{[ \hd_i , \hd_j]} \, A_J  \!\!\!& = & \!\!\!
\dfrac{\e_{ij}}{\th}D_J\dfud{\Act}{\l}
\,, \nonumber \\ 
{[\hd_i, \hd_j ]} \, \l 
 \!\!\!& = & \!\!\!
\dfrac{\e_{ij}}{\th}D_J\dfud{\Act}{A_J}\,,  
\label{new-new} \\
{[ \hd_i , \hd_j]} \, \Phi 
 \!\!\!& = & \!\!\! 0
\qquad \qquad \quad {\rm for} \ \Phi \in \{ A_i, c, \bc, B \} 
\, . 
\nonumber 
\end{eqnarray}

\section{Generalized BRST operator}

We can combine the various symmetry operators defined above into a generalized 
BRST operator that we denote by $\tri$:
\begin{equation}
\label{combined-op}
\tri\equiv s +
\xid 
+ \vare ^i \d_i+\mu^i\dd_i
\qquad 
{\rm with} \ \; 
\xid \equiv \xi^i\partial_i
\, .
\end{equation}
Here, the constant parameters $\xi^i$ and $\mu^i$ have ghost-number $1$ and $\vare^i$ 
has ghost-number $2$. The induced field variations read as
\begin{subequations}\label{trafo-combined}
\begin{align}   
&\tri A_i=D_ic
+\xid A_i
\, ,\\
&\tri A_J=D_Jc+\xid A_J+\mu^iF_{Ji}
\, ,\\
&\tri\l=- \ri g \, {[ \l, c]} + \xid \l+\vare^i \eth{ij}n^j\bc
+\mu^i\eth{ij}D_KF^{jK}
\, ,\\
&\tri c= 
\frac{\ri g}{2} \, {[c,  c] } 
+\xid c+\vare^i A_i
\, ,\\
&\tri\bc=B+\xid \bc
\, ,\\
&\tri B=\xid B+\eid\bc
\, , 
\end{align}
\end{subequations}
and imply
\[
\tri F_{iJ} = - \ri g \, {[ F_{iJ} , c]} + \xid F_{iJ}
- \mu^k D_i F_{kJ}
\, .
\]
Imposing that the parameters $\xi^i,\vare^i$ and $\mu^i$   transform as
\begin{equation}
\label{transpara}
\tri\xi^i=\tri\mu^i=-\vare^i \, , \qquad 
\tri\vare^i=0 \, , 
\end{equation}
we conclude that the operator (\ref{combined-op}) is nilpotent on-shell:
\begin{subequations}\label{nilpoten-op}
\begin{align}
&\tri^2 A_i=\vare^j\eth{ij}\var{\Act}{\l} 
\, ,\\
&\tri^2 A_J=\frac{\mu^i\mu^j}{2}\eth{ij}D_J\var{\Act}{\l}
\, ,\label{nilpoten-op2}\\
&\tri^2\l=\frac{\mu^i\mu^j}{2}\eth{ij}D_J\var{\Act}{A_J}
+\vare^i\eth{ij}\var{\Act}{A_j}-\vare^i\inv{\th^2}D_i\var{\Act}{\l}
\, , \label{nilpoten-op3} \\
&\tri^2 c = \tri^2 \bc = \tri^2 B = 0 \, .
\end{align}
\end{subequations}


\section{Slavnov-Taylor and Ward identities}\label{ST-identities}

The Ward identities corresponding to the various symmetries 
of the action can be gathered into a Slavnov-Taylor (ST) identity 
expressing the invariance of an appropriate total action $\St$ under 
the generalized BRST transformations 
(\ref{trafo-combined}),(\ref{transpara}). 
In this respect, we introduce
an external field $\Phi^*$ (i.e. an antifield in the
terminology of Batalin and Vilkovisky~\cite{Batalin:1981})  
for each field $\Phi \in \{ A_{\mu} ,\l,c \}$ since the latter
transform non-linearly under the BRST variations 
--- see e.g. reference~\cite{Piguet:1995}. We note that the external 
sources $A^{*\mu}$ and $\l ^* $ have  ghost-number $-1$ whereas $c^*$ 
has ghost-number $-2$.   

\subsection{ST identity}

In view of the transformation laws (\ref{trafo-combined}) and (\ref{transpara}), the {\em ST identity} reads as
\begin{eqnarray}
\label{slavnov-id}
0= \slav(\St)
\!\!\!& \equiv  & \!\!\!
\dint d^4x \, 
\{ \dsum{\Phi \in \{ A_{\mu} ,\l,c \}}
{}
\dfud{\St}{\Phi^*}\dfud{\St}{\Phi}
\, + \, \lp B+ \xid \bc \rp \dfud{\St}{\bc} 
\\
\!\!\!&  & \!\!\! \qquad \qquad \qquad 
+ \, \lp \xid B + \eid \bc \rp
\dfud{\St}{B} \}
-\vare^i ( \dpad{\St}{\xi^i} + \dpad{\St}{\mu^i} )
\, . 
\nonumber
\end{eqnarray}
This functional equation is supplemented with the {\em gauge-fixing condition}
\eq
\dfud{\St}{B} = n^i A_i \,.
\eqn{gauge-cond}
By differentiating the ST identity with respect to the field $B$, one finds 
\[
0 = 
\dfud{}{B} \slav(\St) = {\cal G} \St - \xid \, \dfud{\St}{B}
\, , \qquad  
{\rm with} \ \; 
{\cal G} \equiv \dfud{\ }{\bc} + n^i \dfud{\ }{A^{*i}} 
\, ,
\]
i.e., by virtue of (\ref{gauge-cond}), the so-called {\em ghost equation}:
\begin{equation}
\label{ghost-eq}
{\cal G} \St = \xid \, (n^i A_i) 
\, .
\end{equation}
The associated homogeneous equation ${\cal G} \bar S =0$ is solved by functionals $\bar S [ \hAstar^i , \dots ]$ which depend on the variables $A^{*i}$ and $\bc$  through the shifted antifield
\eq
\hAstar^i  \equiv A^{*i}  -n^i\bc
\, .
\eqn{hat-A-star}
Thus, the functional $\St[ A,\l,c,\bc, B \, ;A^*,\l^*,
c^* ;\xi,\mu,\vare] $
which solves both the ghost equation (\ref{ghost-eq})
and the gauge-fixing condition (\ref{gauge-cond})
has the form 
\eq
\St = \dint d^4x\, (B + \xid \bc ) n^i A_i 
\, + \, \bar S[A,\l,c \, ; \hAstar^i, A^{*J}  ,\l^*,c^*; \, \xi,\mu,\vare]
\, ,
\eqn{S-bar}
where the $B$-dependent term ensures the validity of 
condition  (\ref{gauge-cond}).

By substituting expression (\ref{S-bar})
into the ST identity (\ref{slavnov-id}), 
we conclude that the latter equation is satisfied if 
$\bS$ solves the {\em reduced ST identity}
\eq
0 = \BB (\bS) \equiv 
\dsum{ \Phi \in \{ A_{\mu} ,\l,c \} }{} 
\dint d^4x \, 
\dfud{\bS}{{\hat\Phi}^*}\dfud{\bS}{\Phi}
\, - \, \vare^i \, ( \dpad{\bS}{\xi^i} \, + \, \dpad{\bS}{\mu^i} )
\, .
\eqn{reduced-ST}
Here,
${\hat\Phi}^*$ collectively denotes all antifields, but with $A^{*i}$ replaced by the shifted antifield (\ref{hat-A-star}).
Following standard practice~\cite{Piguet:1995}, we introduce
the following notation for the external sources:
\[
\rho^{\mu} \equiv A^{*\mu}  \, , \quad 
\gamma \equiv \l ^*   \, , \quad 
\sigma  \equiv c^* \,,\quad {\hat\rho}^i = \hAstar^i \, .
\]
It can be checked along the usual lines (e.g. see~\cite{Piguet:1995})
that the {\em solution of the reduced ST identity} (\ref{reduced-ST})
is given by 
\begin{eqnarray}
\bS \!\!\! & = & \!\!\! 
\dint d^4x\, \left\{ -\frac{1}{4}F_{\mu\nu} F^{\mu\nu}
+ \frac{\th}{2} \l \e^{ij}F_{ij} \right.
\nonumber \\
&& \qquad \quad 
+ \, \hat\rho^i \left( D_ic+\xid A_i \right)
+\rho^{J} \left( D_Jc+\xid A_J+\mu^iF_{Ji} \right) 
\nonumber \\
&& \qquad \quad 
+ \, \gamma 
\left( -\ri g {[\l ,c] } +\xid \l
+\mu^i \dfrac{\e_{ij}}{\th} D_KF^{jK} \right)
+\s \left( 
\frac{\ri g}{2} \, {[c,c]}+\xid c+\vare^i A_i \right) 
\nonumber \\
&& \qquad \quad  
\left. 
+ \left( \frac{\mu^i\mu^j}{2}\, \frac{\epsilon_{ij}}{\th} 
(D_J \rho^{J} )
+ \vare^i \frac{\epsilon_{ij}}{\th} \, \hat\rho^j
- \vare^i 
\dfrac{1}{2\th^2} \, (D_i\gamma ) \right) \gamma 
\right\} 
\, .
\label{Sbar} 
\end{eqnarray}
Note that 
\[
\bS = S_{\rm inv} + S_{\rm antifields} + S_{\rm quadratic}
\, , 
\]
where $S_{\rm inv}$ is the invariant action introduced in 
(\ref{action}),
$S_{\rm antifields}$ represents the linear coupling of the 
shifted antifields  ${\hat\Phi}^*$
to the generalized BRST transformations (\ref{trafo-combined}a-d)
(the $\bc$-dependent term being omitted)
and $S_{\rm quadratic}$, which is quadratic in the shifted antifields,
reflects  the   contact terms appearing  in
the closure relations (\ref{nilpoten-op}).

\subsection{The antighost and ghost equations}

Differentiating the total action (\ref{S-bar})-(\ref{Sbar}) with respect to the ghost field, one obtains
\[
\dfud{\St}{c} = D_i(\rho^i-n^i\bc) + D_J\rho^J - \ri g[\l,\gamma] 
+ \ri g[c,\s]
+\xid\sigma\,.
\]
By substituting the gauge-fixing condition (\ref{gauge-cond}) in the $n^iA_i$ -
dependent term on the right-hand side, we obtain the 
functional identity
\eq
\dfud{\St}{c} + \ri g\left[\bc,\dfud{\St}{B}\right] + \nid\bc
= D_\mu\rho^\mu -\ri g[\l,\gamma] + \ri g[c,\s]
+\xid\sigma\,,
\eqn{antighost}
which is called the {\em antighost equation}~\cite{Piguet:1995,Piguet:1990}. This equation makes sense as an identity for the action functional since the right-hand side is linear in the quantum fields. Moreover it is local, i.e. not integrated, in space-time. 

Similarly, differentiating the total action with respect to the
antighost field, one obtains the {\em ghost field equation} in functional
form:
\eq
\dfud{\St}{\bc} + \ri g\left[c,\dfud{\St}{B}\right] + \nid c -\xid \dfud{\St}{B}
= - \vare^i \frac{\epsilon_{ij}}{\th}n^j\gamma\,.
\eqn{ghost}
The fact that  both the ghost and the antighost field equations can be cast as
such   local  functional identities is an expression of the ghost freedom of
gauge theories quantized in an axial gauge~\cite{Kummer:1961}.

\subsection{Ward identities}

The Ward identities describing the (non-)invariance of $\St$
under the VSUSY-variations $\d_i$, 
the vectorial symmetry transformations $\hat{\d} _i$ 
and the translations $\partial _i$ can be derived from 
the ST identity (\ref{slavnov-id}) by 
differentiating this identity 
with respect to the corresponding constant ghosts $\vare^i, \, \mu^i$
and $\xi^i$, respectively.

For instance, by differentiating (\ref{slavnov-id}) with respect to $\xi^i$
and by taking the gauge-fixing condition 
(\ref{gauge-cond}) into account,
we obtain the {\em Ward identity for translation symmetry}:
\begin{align}
0=\diff{}{\xi^i}\slav(\St)&=
\dsum{\varphi}{}
\intx \; \partial_i \varphi \, \var{\St}{\varphi}
\, , 
\end{align}
where $\varphi \in \{ A_{\mu} , \lambda, c , \bar c, B;
A^*_{\mu} , \lambda^*, c^* \}$.

By differentiating (\ref{slavnov-id}) with respect to $\vare^i$, we obtain 
\begin{eqnarray}
0 = \diff{}{\vare^i}\slav(\St)
\!\!\!&= &\!\!\! 
-\diff{\St}{\xi^i} - \diff{\St}{\mu^i}
+\intx \Big\{\partial_i\bc \, \var{\St}{B}
+\left(B
+\xid \bc\right)\var{}{\bc} \diff{\St}{\vare^i}
\nonumber \\
\!\!\!&&\!\!\! 
\quad +
\; \dsum{\Phi}{}\left[\left(\var{}{\Phi^*}\diff{\St}{\vare^i}\right)
\var{\St}{\Phi}+\var{\St}{\Phi^*}
\left(\var{}{\Phi}\diff{\St}{\vare^i}\right)\right]\Big\}
\, .
\label{rsti}
\end{eqnarray}
From (\ref{S-bar}) and (\ref{Sbar}), we deduce that 
\begin{subequations}
\begin{align}
\diff{\St}{\vare ^i}&=
\intx \, 
\{ \s A_i  + \eth{ij} \hat{\rho}^j \gamma
+\frac{1}{2\th^2} \, \gamma D_i\gamma \} \label{diff-e}
\\
\diff{\St}{\xi^i}&=
\intx \, \{ - \rho ^{\mu} \partial_iA_\mu
-\gamma \partial_i\l
+\s \partial_i c  \} 
\label{diff-xi}
\\ 
\label{diff-mu}
\diff{\St}{\mu^i}&=
\intx \, \{ F_{iJ}\rho^J+\eth{ij} \, (D_KF^{Kj}) \gamma
+\frac{\e_{ij}}{\th} \mu^j (D_J\rho^J) \gamma \} 
\, .
\end{align}
\end{subequations}
Notice that the right-hand sides of the first two equations are linear in the quantum fields, which is not the case for the third one. Insertion of these expressions into equation (\ref{rsti}) yields a {\em broken Ward identity for VSUSY:} 
\begin{equation}
\label{anomward}
{\cal W}_i \St= \Delta_i 
\, .
\end{equation}
Here, 
\begin{align}
\label{ward22}
{\cal W}_i \St= \intx\, \Big\{&\partial_i\bc \, \var{\St}{B}
+A_i \, \var{\St}{c}
+\left( \eth{ij}\left(n^j\bc-\rho^j\right)
+\dfrac{1}{\th^2}
D_i\gamma\right)\var{\St}{\l}
\nonumber\\
& \qquad \qquad \qquad \qquad 
\ + \gamma \eth{ij} \, \var{\St}{A_j}
+\left(\s+ \frac{\ri g}{\th^2}\gamma\gamma\right)\var{\St}{\rho^i}\Big\}
\end{align}
and 
\begin{equation}
\label{wardid2}
\Delta_i =
\diff{\St}{\xi^i}+\diff{\St}{\mu^i}
+ \intx\, 
\eth{ij}n^j\left(B+\xid \bc\right)\gamma
\, . 
\end{equation}
More explicitly, 
$\Delta_i = \Delta_i \Big|_{\xi=\mu=0} + B_i [\xi , \mu]$ with 
\begin{eqnarray}
\Delta_i \Big|_{\xi=\mu=0} 
\!\!\! &=& \!\!\!
\intx \, 
\{ -\rho^{\mu} \partial_i A_{\mu} +\s\partial_ic
-\gamma\partial_i\l+\gamma\eth{ij}n^jB-\rho^J  F_{Ji}
- \gamma \frac{\e_{ij}}{\th}D_K F^{jK} \}
\nonumber \\
B_i [\xi , \mu] 
\!\!\! &=& \!\!\!
\intx \, \{ \xid \bc \, \eth{ij} n^j \gamma 
+ \eth{ij} \mu^j (D_J \rho^J) \gamma \}
\, .
\label{break}
\end{eqnarray}
Several remarks concerning the results (\ref{anomward})-(\ref{break}) are in order.
First, we note that the field variations given by (\ref{ward22}) extend the VSUSY transformations (\ref{susy}) by source dependent terms.
It is the presence of the sources which leads to a breaking $\Delta_i$ of VSUSY --- cf. the unbroken Ward identity (\ref{ward-susy}) for the gauge-fixed action.
Second, we remark that the breaking of VSUSY is \emph{non-linear} in the quantum fields: The non-linear contributions are contained in $\Delta_i \Big|_{\xi=\mu=0}$
and given by 
\[
- \intx \, 
\{ \rho^J  F_{Ji} + \gamma \frac{\e_{ij}}{\th} D_K F^{jK} \}
= 
- \intx \, 
\{ \rho^J (\hat\d_iA_J) + \gamma (\hat\d_i\l) \}
\, , 
\]
where $\hat\d_i$ are the  vectorial symmetry transformations (\ref{newsymmetry}).  However, these non-linear breakings (which could jeopardize a non-ambiguous definition of the theory) are contained in the derivative $\partial\St/\partial\mu^i$ and are therefore functionally well defined.

Finally, we come to the Ward identity for the  vectorial symmetry $\hat\d_i$. By differentiating the ST identity (\ref{slavnov-id}) with respect to $\mu^i$ and using (\ref{diff-mu}), one finds \begin{align}
\intx\Bigg\{&-F_{iJ} \, \var{\St}{A_J}
-\frac{\e_{ij}}{\th}\left(D_KF^{Kj}
+\mu^j \, D_K\rho^K\right)\var{\St}{\l}
+D_K\rho^K \, 
\var{\St}{\rho^i}
\nonumber\\
& + \eth{ij}D_KD^K\gamma \, \var{\St}{\rho_j}
-\left( D_i\rho^I+\eth{ij}D^jD^I \gamma 
+\ri g\eth{ij} \co{F^{Ij}}{\gamma}\right)\var{\St}{\rho^I}
\nonumber\\
&+ \ri g \eth{ij} \mu^j \co{\rho^I}{\gamma}\var{\St}{\rho^I}\Bigg\}
\, = -\intx\, \eth{ij} \, \vare ^j (D_K\rho^K) \gamma
\, ,
\end{align}
i.e. we have  here a breaking which is linear in the quantum fields.

\section{Consequences of VSUSY}\label{loop-consequences}
The generating functional $Z^c$ of the connected Green functions is given by the Legendre  transform\footnote{ In the ``classical approximation'', the generating functional $\Gamma$ of the one-particle-irreducible Green functions is equal to the total classical action $\St$. Its Legendre transform $Z^c$ generates the
connected Green functions in the tree graph approximation.}
\begin{align}
 Z^c[j_A,j_\l,j_B,j_c,j_\bc]
=\Gam[A,\l,B,c,\bc]
+\int d^4x\left(j_A^\mu A_\mu
+j_\l\l +j_BB
+j_c c+j_\bc\bc\right)
\, .
\end{align}
Thus, we have the usual relations
\begin{align}
&\var{Z^c}{j_A^\mu}=A_\mu \, ,
&&\var{Z^c}{j_\l}=\l \, , 
&&\var{Z^c}{j_B}=B \, ,
&& \var{Z^c}{j_c }=c \,,
&&\var{Z^c}{j_\bc}=\bc \,,
\nonumber\\
&\var{\Gam}{A_\mu}=-j_A^\mu \, ,
&&\var{\Gam}{\l}=-j_\l \, ,
&&\var{\Gam}{B}=-j_B \, ,
&&\var{\Gam}{c}=j_c  \, ,
&&\var{\Gam}{\bc}=j_\bc \, , 
\end{align}
and 
\begin{align}
& \var{Z^c}{\Phi^*}=\var{\Gamma}{\Phi^*}\,, 
&&\dpad{Z^c}{\xi^i}=\dpad{\Gamma}{\xi^i}\,,
&&\dpad{Z^c}{\vare^i}=\dpad{\Gamma}{\vare^i}\,,
&&\dpad{Z^c}{\mu^i}=\dpad{\Gamma}{\mu^i} \; .
\end{align}

For vanishing antifields, the {\em Ward identity} describing the VSUSY (\ref{anomward}) becomes in terms of $Z^c$: 
\begin{align}\label{ward2}
\mathcal{W}_iZ^c
=\int d^4x \, \Big\{ & 
j_B \, \partial_i \var{Z^c}{j_\bc}   
- j_c \, \var{Z^c}{j_A^i} \, 
+ \eth{ij} n^j  j_\l \,  \var{Z^c}{j_\bc} \,  \Big\}=0 \, .
\end{align}
By varying (\ref{ward2}) with respect to the appropriate sources, one gets the following relations  for the two-point functions (i.e. the free propagators):
\begin{equation}
\label{rel-all}
\vvar{Z^c}{j_A^i}{j_\l}\Bigg|_{j=0}
=-\eth{ij} n^j\vvar{Z^c}{j_\bc}{j_c }\Bigg|_{j=0}
\, , \qquad \quad 
\vvar{Z^c}{j_A^i}{j_A^\nu}\Bigg|_{j=0}=0
\, .
\end{equation}
The {\em gauge-fixing condition} (\ref{gauge-cond}) 
is equivalent to $n^i\, \var{Z^c}{j_A^i} = -j_B$,
from which it follows that 
\begin{align}\label{rel-AB-prop}
n^i\vvar{Z^c}{j_B(y)}{j_A^i(x)}\Bigg|_{j=0}
=-\d ^{(4)}(x-y) \, .
\end{align}
For vanishing antifields, 
the {\em antighost equation} (\ref{antighost}) 
can be written as
\[
-n \cdot \partial\var{Z^c}{j_\bc }- \ri g
\left[ j_B , \var{Z^c}{j_\bc } \right]
=j_c \, ,
\]
and by varying this equation with respect to $j_c$, one concludes that \begin{align}\label{rel-ghost-eq}
n \cdot \partial\vvar{Z^c}{j_c(x)}{j_\bc (y)}\Bigg|_{j=0}=-\d ^{(4)}(x-y) \, .
\end{align}
Note that the same result may be obtained from the {\em ghost equation} (\ref{ghost}) which reads  
in terms of $Z^c$ (for vanishing antifields and $\xi^i=0$):
\begin{align}\label{ghost-sol}
-n \cdot \partial\var{Z^c}{j_c } - \ri g
\left[ j_B , \var{Z^c}{j_c } \right]
=j_\bc \, .
\end{align}

In momentum space, the free propagators of the theory
are given by
\begin{subequations}\label{propagators}
\begin{align}
&\ri \Delta^{c\bc}(k)=-\inv{nk}
 \, ,\qquad 
\ri \Delta^{AB}_\mu(k)=\frac{\ri k_\mu}{nk} \, ,\\
&\ri \Delta_\mu^{A\l}(k)=\inv{\k^2}
\left(\k_\mu-k_\mu\frac{n\k}{nk}\right) \, ,\\
&\ri \Delta_{\mu\nu}^A(k)
=\frac{-\ri}{k^2}\left[g_{\mu\nu}
-\frac{n_\mu k_\nu+n_\nu k_\mu}{nk}+a\frac{k_\mu k_\nu}{(nk)^2}
+b ( k_\mu\tilde k_\nu
+ k_\nu\tilde k_\mu )
-\frac{\tilde k_\mu\tilde k_\nu}{\tilde k^2}\right] \, ,
\end{align}
\end{subequations}
with $(g_{\mu \nu}) = {\rm diag} \, (1,-1,-1,-1)$ and~\footnote{
We have $\k^2=-\th^2(k_1^2+k_2^2)$, $nk=-(n_1k_1+n_2k_2)$, $n\k=\th(n_1k_2-n_2k_1)$
and $\ri \Delta^{AB}_\mu(x-y)=-\ri \vvar{Z^c}{j_B(y)}{j_A^\mu(x)}\Big|_{j=0}$.}
\begin{align}
&\k_i\equiv\th\epsilon_{ij}k^j  
 \, ,\qquad
\k_J\equiv 0 \, , \nonumber\\
&a\equiv n^2-\frac{(n \tilde k)^2}{\tilde k^2}
 \, ,\qquad b\equiv \frac{n\tilde k}{(nk)\tilde k^2} \, .
\end{align}
One can easily check that these propagators obey the conditions (\ref{rel-all}), (\ref{rel-AB-prop}) and (\ref{rel-ghost-eq}).

\begin{figure}[h]
\centering
\includegraphics[scale=0.9]{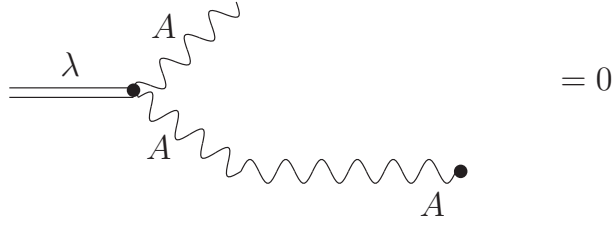}
\caption{The $\l AA$-vertex contracted with a photon propagator vanishes.}
\label{fig:vertex-prop}
\end{figure}

\begin{figure}[h]
\centering
\includegraphics[scale=0.9]{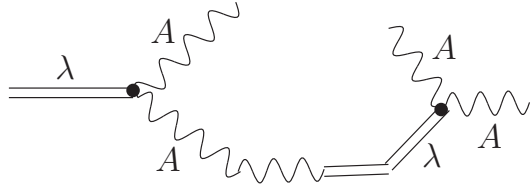}
\caption{Building a Feynman loop graph with a $\l AA$-vertex is impossible 
without a photon propagator.}
\label{fig:impossible}
\end{figure}

\begin{figure}[h]
\centering
\includegraphics[scale=0.6]{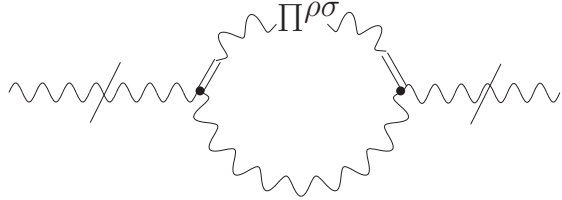}
\caption{The ``problematic'' 2-loop graph vanishes in this case.}
\label{fig:2-loop}
\end{figure}

As we are now going to show, the remarkable outcome of the identities (\ref{rel-all}), (\ref{rel-AB-prop}) and (\ref{rel-ghost-eq}) is that they are sufficient for killing all possible IR divergences in the radiative corrections. The second relation in (\ref{rel-all}), which states that the photon propagators  $\Delta_{i\nu}^A$ vanish, has an important consequence. Indeed, since the $\l AA$-vertex is proportional to $\theta^{ij}$, all Feynman graphs which include a $\l AA$-vertex contracted with an internal photon line \emph{must cancel} (cf. Figure~\ref{fig:vertex-prop}).
But since it is obviously impossible to construct a Feynman graph 
(except for a tree graph) including $\l AA$-vertices which do not couple 
to internal photon propagators, {\em all loop corrections involving 
the $\l AA$-vertex have to vanish}! 
Note, that a mixed photon-$\l$ propagator contracted with a $\l AA$-vertex 
leads to the necessity of another $\l AA$-vertex, 
and so in order to build a closed loop, photon propagators are necessary 
(see Figure~\ref{fig:impossible}).
Hence, the Feynman rules involving the $\lambda$-field do not enter the loop corrections of the photon $n$-point function. In particular, the IR-problematic graph mentioned in our previous paper~\cite{Blaschke:2005c} and depicted in Figure~\ref{fig:2-loop} is absent for our choice of gauge.
Now that we have shown that the $\l$-field plays no role in the radiative corrections of the gauge field, the absence of IR-divergences follows from the line of arguments 
given in reference ~\cite{Slavnov:2004}.

From these considerations, it should also become obvious that all loop corrections 
to the $\lambda$-propagator and the mixed $\lambda$-photon propagator vanish, leaving the tree approximation as the exact solution for this sector. Furthermore, equations (\ref{rel-AB-prop}) and (\ref{rel-ghost-eq}) provide exact solutions to the AB propagator and the ghost propagator~\cite{Schweda:1996}. Also notice that the first of equations (\ref{rel-all}) 
is consistent with the considerations above: it gives us the exact solution for the mixed $\lambda$-photon propagator once the solution for the ghost propagator is found from (\ref{rel-ghost-eq}).

\section{Conclusion}
As discussed in section \ref{vector-susy}, the
$U(1)$-NCGFT with Slavnov term and with an appropriate axial gauge-fixing  
exhibits a far richer symmetry structure than initially expected. In
particular, it admits a linear 
VSUSY which is similar to the one present 
in the $2$-dimensional $BF$ model, provided one chooses the deformation matrix
$\th^{\mu\nu}$ to be space-like and the axial gauge-fixing vector
$n^\mu$ to lie in the plane of the noncommuting coordinates.
While this VSUSY yields a superalgebra (which includes the BRST
operator $s$ and the translation generator in the noncommutative plane), 
it differs from the one present in the noncommutative 
$2$-dimensional $BF$ model by the fact that it contains an additional
nonlinear vectorial symmetry (given by the transformation laws
(\ref{newsymmetry})).

As a consequence  of the identities for the free propagators which follow
from the VSUSY, all loop corrections become independent 
of the $\l AA$-vertex.
This is the reason why the theory 
in our particular space-like axial gauge is finite,
as pointed out by Slavnov in reference~\cite{Slavnov:2004}. 

Thus, the absence of IR singularities in a NCGFT can be achieved by other
means than extending it to a Poincar\'e supersymmetric
theory\footnote{  The role of Poincar\'e supersymmetry for the
cancellation of IR singularities has been extensively studied in the
literature --- see~\cite{Gomes:2004} for a review and further references.
 }
(as was already emphasized by
Slavnov~\cite{Slavnov:2004}), namely by modifying it physically by
adding the Slavnov term (which leads to the presence of VSUSY
that is characteristic for a class of gauge-fixings). One may note that a
supersymmetry is again responsible for the cancellation of IR
singularities. But, contrary to the Poincar\'e supersymmetry which
is physical, VSUSY is not physical, its existence following from the
specific choice we have made for the gauge-fixing\footnote{
It has been shown~\cite{Viena-gomes} that Chern-Simons models 
without Poincar\'e supersymmetry  may also be
free of the IR singularities, depending on the gauge-fixing choice and on the 
coupling with matter.}.  
\subsection*{Acknowledgments}
Olivier Piguet would like to thank the Institut de Physique Nucl\'eaire of the University of Lyon for a financial help, which permitted a stay during which a substantial part of this work has been done. Fran\c cois Gieres kindly acknowledges his stay at the University of Vit\'oria at the final stage of the present work.



\end{document}